\documentclass[12pt,a4paper]{article}

\usepackage{graphicx}
\usepackage{amssymb}
\usepackage{bm}
\newcommand{\be}{\begin{equation}}
\newcommand{\ee}{\end{equation}}
\newcommand{\bel}[1]{\be\label{#1}}
\newcommand{\re}[1]{Eq.~(\ref{#1})}
\newcommand{\mbs}[1]{\mbox{$\scriptstyle{#1}$}}
\newcommand{\psib}{\overline{\psi}}
\newcommand{\dd}{\partial\hspace{-6pt}/}
\newcommand{\ds}{\displaystyle}
\newcommand{\hsp}{\hspace*{1pt}}

\begin{document}

\begin{center}
{\Large\bf  Constraints on possible phase transitions\\
above the nuclear saturation density}\\[5mm]
{\bf I.N.~Mishustin$^{\,1,2,3}$, L.M.~Satarov$^{\,1,2}$, H.
St\"ocker$^{\,1}$, and W. Greiner$^{\,1}$}
\end{center}
\begin{tabbing}
\hspace*{1.5cm}\=
${}^1$\={\it Institut~f\"{u}r~Theoretische~Physik,
J.W.~Goethe~Universit\"{a}t,}\\
\>\>{\it D--60054~Frankfurt~am~Main,~\mbox{Germany}}\\
\>${}^2$\>{\it The Kurchatov~Institute, Russian Research Center,}\\
\>\>{\it 123182~Moscow,~\mbox{Russia}}\\
\>${}^3$\>{\it The Niels~Bohr~Institute,
DK--2100~Copenhagen {\O},~\mbox{Denmark}}\\
\end{tabbing}

\begin{abstract}

We compare different models for hadronic and quark phases of cold
baryon--rich matter in an attempt to find a deconfinement phase
transition between them. For the hadronic phase we consider
Walecka--type mean--field models which describe well the nuclear
saturation properties. We also
use the variational chain model which takes into account correlation
effects. For the quark phase we consider the MIT bag model, the
Nambu--Jona-Lasinio and the massive quasiparticle models.  By comparing
pressure as a function of baryon chemical potential we find that
crossings of hadronic and quark branches are possible only in some
exceptional cases while for most realistic parameter sets these
branches do not cross at all. Moreover, the chiral phase transition,
often discussed within the framework of QCD motivated models, lies
in the region where the quark phases are unstable with respect to the
hadronic phase. We discuss possible physical consequences of these
findings.

\end{abstract}

\baselineskip 24pt



\section{Introduction}
It is commonly believed that quarks and gluons are relevant
degrees of freedom in strongly interacting matter at very high
temperatures and baryon densities. This state of matter is usually
described by various QCD motivated models. On the other hand, at low
temperatures and moderate densities, at least up to the nuclear
saturation density $\rho_0=0.17$~fm$^{-3}$, strongly interacting
matter exists in the hadronic phase. In particular, atomic nuclei
are finite droplets of this phase with baryon density $\rho_B\simeq\rho_0$.
They are self--bound and therefore can exist in vacuum, without
external pressure. This fact itself provides an important constraint
on the equation of state (EOS) of cold nuclear matter, namely, its
pressure must vanish at $\rho_B=\rho_0$. There are many effective
models which successfully describe nuclear matter in terms of
interacting nucleons.

Unfortunately, at present there exists no rigorous approach which
can describe the EOS of strongly interacting matter at finite baryon
densities. As known, QCD lattice simulations have principal limitations
at nonzero chemical potential. Therefore the only practical way
to study the possibility of the deconfinement phase transition in this
case is to compare various models of hadronic and quark phases.
In the last two decades there were numerous attempts to
construct a unified EOS which would interpolate between the two
asymptotic regimes. In this paper we critically revise this
problem in the light of new calculations for the hadronic and
quark phases.

In our analysis we include four hadronic models. Two of them are
Relativistic Mean--Field (RMF) models of the Walecka
type~\cite{Ser85,Ser97}, namely, the NLZ~\cite{Ruf89} and the
TM1~\cite{Sug94} models. Next is the so called Chiral
Hadronic Model (CHM) which was recently developed in
Refs.~\cite{Pap98,Pap99}. Fourth model is the Variational Chain
Model~(VCM)~\cite{VCM98} based on the Argonne NN potentials with
addition of 3--body forces and relativistic corrections. Unlike the
above three models, the VCM takes into account correlation effects
neglected in the mean--field approximation. As argued in
Ref.~\cite{VCM98}, the mean--field approximation is not well justified
at baryon densities~$\rho_B\lesssim\rho_0$\,.

For the quark matter we take three different models: the MIT Bag
Model~(BM)~\cite{Cho74}, the Massive Quasiparticle
Model~(MQM)~\cite{Pes96,Sch97} and the Nambu--Jona-Lasinio
model~(NJL)~\mbox{\cite{Nam61,Lar61}}. Presumably, the hadronic degrees of
freedom are more relevant at baryon densities $\rho_B\lesssim\rho_0$
and quark degrees of freedom take over at much larger densities. But
the transition between these two regimes is very poorly understood at
present. By considering various hadronic and quark models we pursue
several goals. First, we compare different models for a single phase
to get an idea on the uncertainty in their predictions. Second, by
applying the Gibbs criterion, we investigate the possibility of a
deconfinement phase transition in cold baryon-rich matter. Finally,
we examine reliability of different quark models (in particular,
allowed values of model parameters) by extrapolating their predictions
into the domain of nuclear matter, $\rho_B\sim\rho_0$.

The paper is organized as follows. In Sect. II we give short
descriptions of several popular models of the hadronic phase.
Their predictions regarding the equation of state of cold baryonic
matter are summarized and compared between each other.In Sect. III
different models of the quark phase are introduced. Their pairwise
comparison with hadronic models, aimed at finding a hadron--quark phase
transition, is carried out in a systematic way. In Sect. IV we present
our conclusions and outlook.

\section{Models of hadronic phase}

\subsection{Relativistic mean--field models}\label{sec:rmf}

At present the field--theoretical description of dense
hadronic matter is one of the most popular approaches.
Within this approach the matter is described in terms of baryons
interacting with self--consistent meson fields. Most calculations are
done within the mean--field approximation. There are many
versions of the RMF model which differ by the choice of meson fields
as well as by the baryon--meson coupling schemes. Here
we consider two realizations of the RMF approach which give
very good description of finite nuclei, namely, the NLZ~\cite{Ruf89}
and the TM1~\cite{Sug94} models.

The general form of the effective RMF Lagrangian used in these models is\,%
\footnote{
Below we consider static and
homogeneous isospin--symmetric matter. In this case the derivatives of
mean meson fields over space and time as well as the contribution of
$\rho$--mesons may be omitted.
}
\bel{lagr}
{\cal L}=\psib\left(i\hsp\dd-m_N^*-
g_\omega\omega\gamma_0\right)\psi-U_s(\sigma)+U_v(\omega)\,.
\ee
Here $\psi, \sigma$ and $\omega$ are, respectively, the nucleon,
scalar and vector meson fields,
\bel{nmas}
m_N^*=m_N-g_\sigma\sigma
\ee
is the effective nucleon mass, $U_s$ and $U_v$ are the
scalar and vector potentials:
\begin{eqnarray}
U_s(\sigma)&=&\frac{(m_\sigma\sigma)^2}{2}
+\frac{g_2\hsp\sigma^3}{2}+\frac{g_3\hsp\sigma^4}{4}\,,\label{spot}\\
U_v(\omega)&=&\frac{(m_\omega\hsp\omega)^2}{2}+
\frac{g_{3\omega}\hsp\omega^4}{4}\,.\label{vpot}
\end{eqnarray}
In the above equations $m_i$ denote vacuum masses of nucleons ($i=N$)
and mesons ($i=\sigma,\omega$), $g_j$~are coupling constants. The
parameter sets for the NLZ and TM1 models
are listed in Table~\ref{tab1}. Note that contrary to the NLZ, within the
TM1 model the vector field is a nonlinear function of the baryon
density. This leads to a significant reduction of the repulsive interaction
at high $\rho_B$.
\begin{table}[h]
\caption{Parameters of RFM models}
\label{tab1}
\vspace*{5mm}
\hspace*{-5mm}
\begin{tabular}{|c|c|c|c|c|c|c|c|c|}\hline\hline
\hspace*{1.5cm}&\mbox{$m_N$\hsp(GeV)}&\mbox{$m_\sigma$\hsp(GeV)}&
\mbox{$m_\omega$\hsp(GeV)}&$g_\sigma$&\mbox{$g_2$ (fm$^{-1}$)}&$g_3$&%
$g_\omega$&$g_{3\omega}$\\
\hline
NLZ~\cite{Ruf89}& 938.9 & 488.67 & 780 & 10.0553 & -13.5072 & -40.2243 & 12.9086 & 0 \\
TM1~\cite{Sug94}& 938.0 & 511.198 & 783 & 10.0289 & -7.2325 & 0.6183 & 12.6139 & 71.305\\
\hline
\end{tabular}
\end{table}

Within the mean--field approximation $\sigma$ and $\omega$ fields are
regarded as purely classical and replaced by c--numbers. Using the
Lagrangian (\ref{lagr}) one can easily calculate pressure $P$ of cold
nonstrange matter as a function of the baryon chemical potential $\mu_B$
(see for details Refs.~\cite{Ser85,Ser97}). At given $\mu_B$\,,
applying thermodynamic relations one obtains the following equations
for the baryon density and the energy density of matter
\begin{eqnarray}
\rho_B&=&dP/d\mu_B\,,\label{dens}\\
\epsilon&=&\mu_B\rho_B-P\,.\label{enden}
\end{eqnarray}
The energy per baryon is equal to $E/B=\epsilon/\rho_B$\,.

\subsection{The chiral hadron model}\label{sec:chm}

Although the Lagrangian~(\ref{lagr}) leads to very good description
of nuclear phenomenology it has one principal defect. Namely,
it does not respect chiral symmetry of strong interaction. It is commonly
accepted that this symmetry is spontaneously broken in vacuum and can
be restored at high density and temperature. In recent years there were
several attempts to incorporate this symmetry into the RMF framework.
It turned out that most easily this can be achieved~\cite {Ell94,Mis93}
by introducing an
additional scalar (dilaton) field $\chi$ responsible for the trace anomaly
of QCD. For our analysis we have chosen the Chiral Hadron Model (CHM)
developed in Ref.~\cite{Pap98,Pap99}. Below we use the version of the CHM,
given by the parameter set C1 in Ref.~\cite{Pap99}. As compared to the
NLZ and TM1 models, the CHM includes also a strange scalar field $\zeta$.
The following parametrization of $m_N^*$ is used instead of (\ref{nmas}):
\bel{nmas1}
m_N^*=g_\sigma\sigma+g_\zeta\,.
\ee
The scalar potential $U_s$ is parametrized as a fourth-order polynomial in
$\sigma,\,\zeta$ and $\chi$. Its parameters are tuned to reproduce
the pattern of spontaneous symmetry breaking at low densities. In addition
there appear logarithmic terms, $\propto\chi^4\log{\chi^4}$ and
$\propto\chi^4\log{(\sigma^2\zeta)}$, motivated by the trace anomaly.
As shown in Ref.~\cite{Pap99} the CHM gives satisfactory
description of nuclear matter and finite nuclei.

\subsection{The variational chain model}\label{sec:vcm}

The VCM~\cite{VCM98} is a microscopic approach where the interaction
between nucleons is describes in terms of two and three body
forces. The Hamiltonian is written in the form
\bel{vcmh}
{\cal H}=-\sum_{i}\frac{1}{4}
\left(\frac{1}{m_p}+\frac{1}{m_n}\right)\Delta_i+
\sum_{i<j}(v_{ij}+\delta v_{ij})+\sum_{i<j<k}v_{ijk}\,.
\ee
Here the first term gives non--relativistic kinetic energy of nucleons.
The NN interaction terms~$v_{ij}$ are chosen in the form of the Argonne
potential with parameters fitted to reproduce the NN scattering phases
up to the c.m. energy 300 MeV. The terms $\delta v_{ij}$ denote
relativistic (Lorentz--boost) corrections up to the second order in the
NN pair momentum. The 3N interaction terms $v_{ijk}$ are taken in the
Urbana UIX form with parameters which correctly reproduce binding energies
of lightest ($A\leq 4$) nuclei. The calculations are carried out by
applying the variational Monte Carlo method and the chain summation
technique~\cite{VCM98}. Besides properties of finite nuclei,
pressure and energy density of nuclear matter are calculated at
various nucleon densities and neutron to proton ratios.

\subsection{Comparison of model predictions}\label{sec:cmp}

\begin{figure}
\centerline{
\includegraphics[height=9cm]{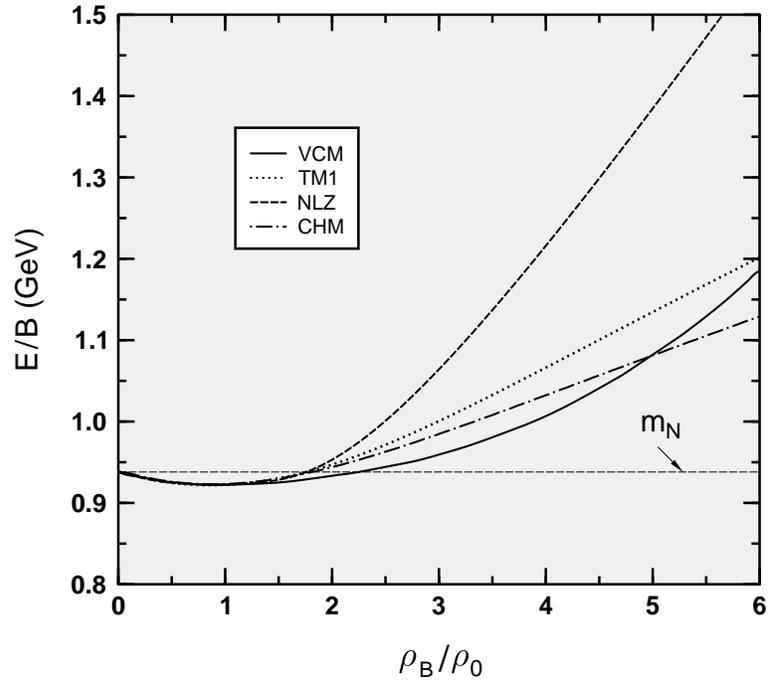}
}
\caption
{Energy per baryon  of cold nuclear matter calculated within different
hadronic models as function of baryon density.
Here and below the curves are specified in the key box.}
\label{fig1}
\end{figure}

\begin{figure}
\centerline{
\includegraphics[height=9cm]{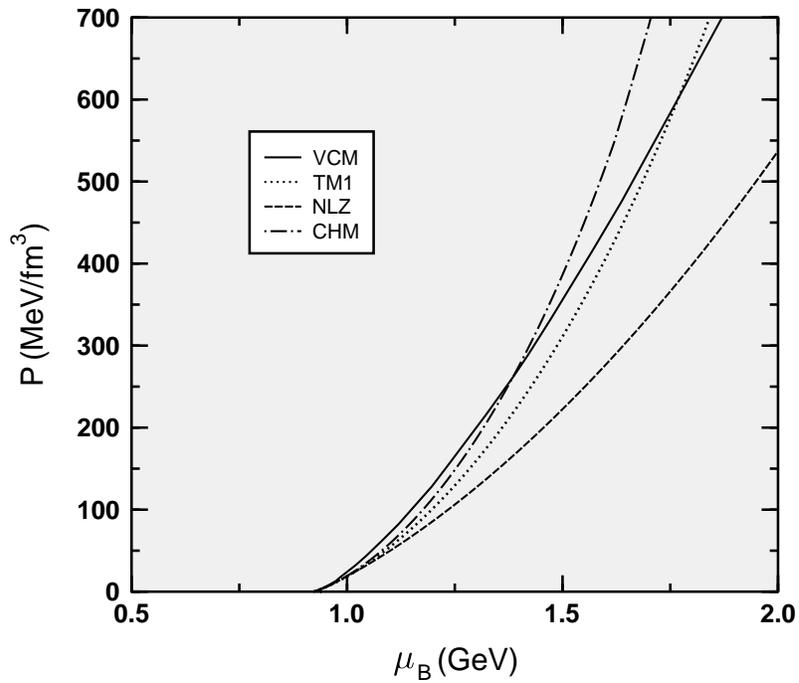}
}
\caption
{Pressure of cold nuclear matter vs baryon chemical
potential calculated within different hadronic models.}
\label{fig2}
\end{figure}

\begin{figure}
\centerline{
\includegraphics[height=9cm]{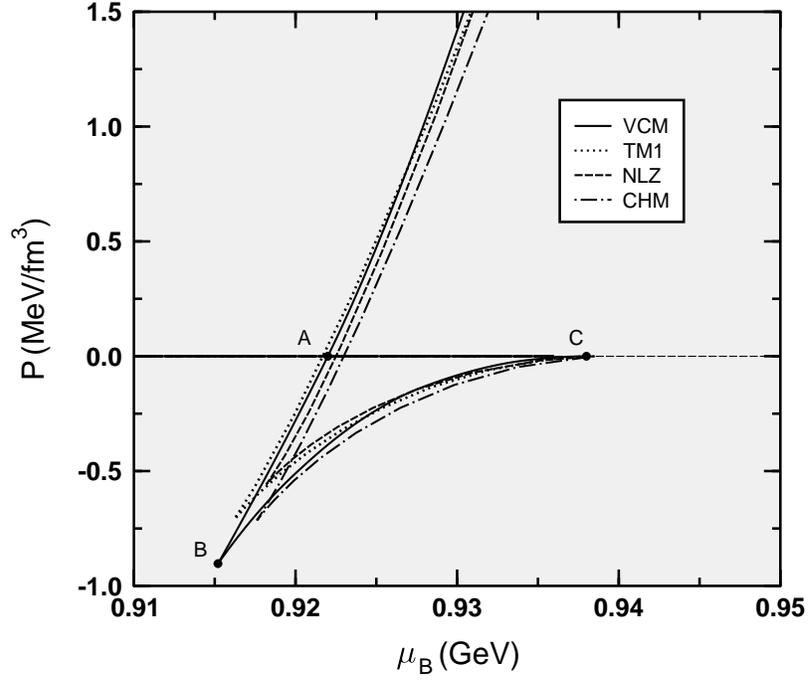}
}
\caption
{The same as in Fig.~\ref{fig2}, but for small values of baryon
chemical potential. Points A,B, and C show boundaries of metastable and
unstable regions in the case of the VCM.}
\label{fig3}
\end{figure}

\begin{figure}
\centerline{
\includegraphics[height=9cm]{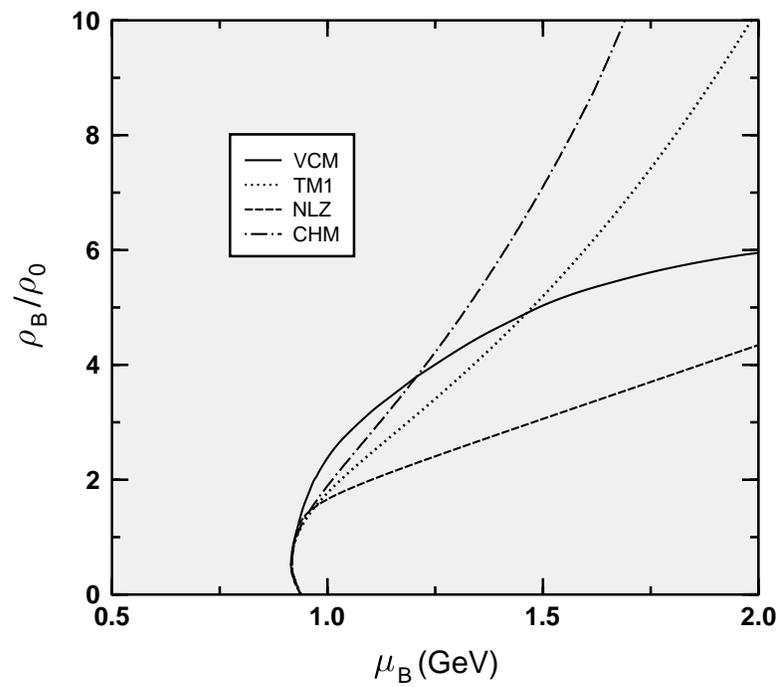}
}
\caption
{Baryon density of cold nuclear matter as
function of chemical potential.}
\label{fig4}
\end{figure}

\begin{figure}
\centerline{
\includegraphics[height=9cm]{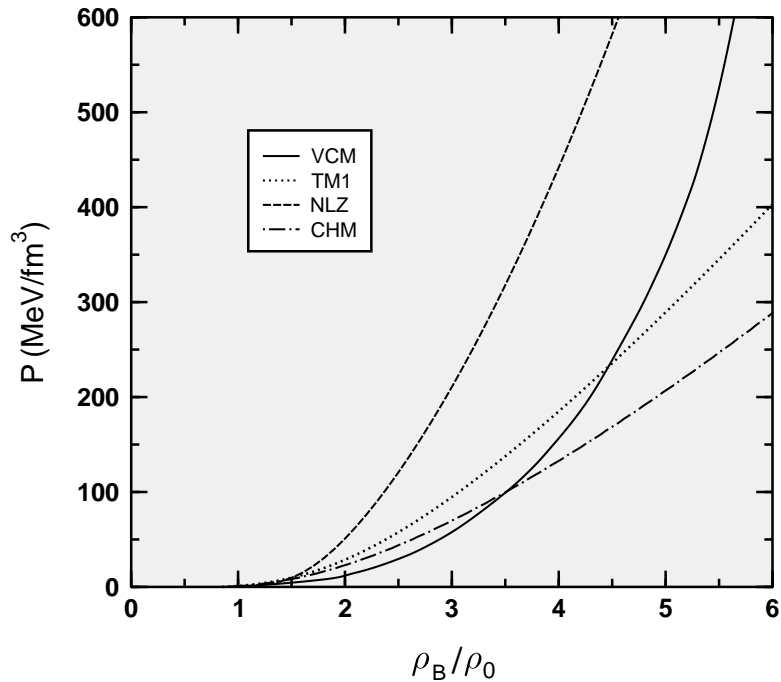}
}
\caption
{Pressure vs baryon density calculated within different hadronic models.}
\label{fig5}
\end{figure}

In Figs.~\ref{fig1}--\ref{fig5} we compare pressure and binding
energy of cold isospin--symmetric baryonic matter calculated within
the above four models. Figure~\ref{fig1} shows predictions for
energy per baryon as a function of $\rho_B$\,. All models give
nearly the same results at densities $\rho_B\lesssim\rho_0$\,. At
higher densities the CHM, VCM, and the TM1 model are not far from each
other, unlike the NLZ model which seems to overestimate significantly
the energy per baryon at $\rho_B\gtrsim 3 \rho_0$\,%
\footnote{
 One should regard the results of
 effective hadronic models at $\rho_B>>\rho_0$ with caution due to
 the lack of available information concerning EOS at such
 densities.
}.

Figure~\ref{fig2} shows pressure as a function of baryon chemical
potential. The discrepancy between different hadronic models becomes
especially evident at $\mu_B-m_N\gtrsim 200$ MeV. One can see that the
results of the NLZ model strongly deviate from the predictions of other
models. Apparently, this is due to overestimation of vector repulsion
within the NLZ model. On the other hand, the CHM, VCM, and the TM1
model give similar results at $\mu_B\lesssim 1.5$ GeV. All hadronic
models predict the first order phase transition of the liquid--gas type
at $\mu_B\simeq m_N$ which corresponds to baryon densities
$\rho_B\lesssim\rho_0$. This is clear from Fig.~\ref{fig3} where
pressure is shown at a finer scale. For example, in the case of the VCM
the parts AB, BC, and AC of the pressure curve are unstable with
respect to decomposition of matter into the dilute ($i=1$) and dense
($i=2$) phases. According to the Gibbs rules for coexisting phases
\begin{eqnarray}
P^{(1)}&=&P^{(2)}\,,\label{gr1}\\
\mu_B^{(1)}&=&\mu_B^{(2)}\,.\label{gr2}
\end{eqnarray}
The phase transition points at $T=0$ are given by intersection of
different branches of pressure as a function of chemical potential\,%
\footnote{\label{fpag}
 When analyzing EOS in the $\mu_B-P$ plane one should bear in mind
that lower branches of pressure correspond to metastable or unstable
states.
}.
For example, point A in Fig.~\ref{fig3} is the phase
transition point in the case of the VCM.

According to \re{dens}, the density jump in the first order phase
transition, $\rho_B^{(2)}-\rho_B^{(1)}$ is equal to the difference of
slopes $dP/d\mu_B$ at $\mu_B\to\mu_c\pm 0$ where $\mu_c$ (the phase
transition point) is the solution of \re{gr1}. The equivalent criterion
of the first order phase transition is the so--called ''double tangent
construction'' for energy density as a function of baryon density.
Indeed, by using thermodynamic identities~(\ref{dens})--(\ref{enden})
one can rewrite Eqs.~(\ref{gr1})--(\ref{gr2}) in the form
\bel{dtcn}
\frac{\ds\epsilon^{(2)}-\epsilon^{(1)}}
{\ds\rho_B^{(2)}-\rho_B^{(1)}}=\frac{d\epsilon^{(1)}}{\ds d\rho_B}
=\frac{d\epsilon^{(2)}}{\ds d\rho_B}\,.
\ee
Below the conditions~(\ref{gr1})--(\ref{gr2}) and (\ref{dtcn})
will be used for investigating the possibility and estimating parameters of
the deconfinement transition by comparing pressures of hadronic ($i=1$)
and quark~($i=2$) phases.

Four hadronic models considered in this paper predict essentially
different behavior of the baryon density with raising~$\mu_B$\,. This
is shown in Fig.~\ref{fig4}. Again, one can see strong deviation of
the NLZ results from the predictions of other hadronic models.
Figure~\ref{fig5} shows pressure isotherms as functions of~$\rho_B$. It
is clearly seen that the NLZ model predicts the ''hardest'' EOS.
Indeed, at intermediate  densities $\rho_B\sim (1.5-4)\,\rho_0$ this
model predicts the highest compressibility i.e. the largest derivative
$dP/d\rho_B$. The NLZ model gives also the largest pressure at given
$\rho_B$\,. On the other hand, the $P(\mu_B)$ curve predicted by this
model is lower as compared to other models (see Fig.~\ref{fig2}).
Apparently, this difference is caused by the vector repulsion which is
especially large in the NLZ model.

\begin{figure}
\centerline{
\includegraphics[height=9cm]{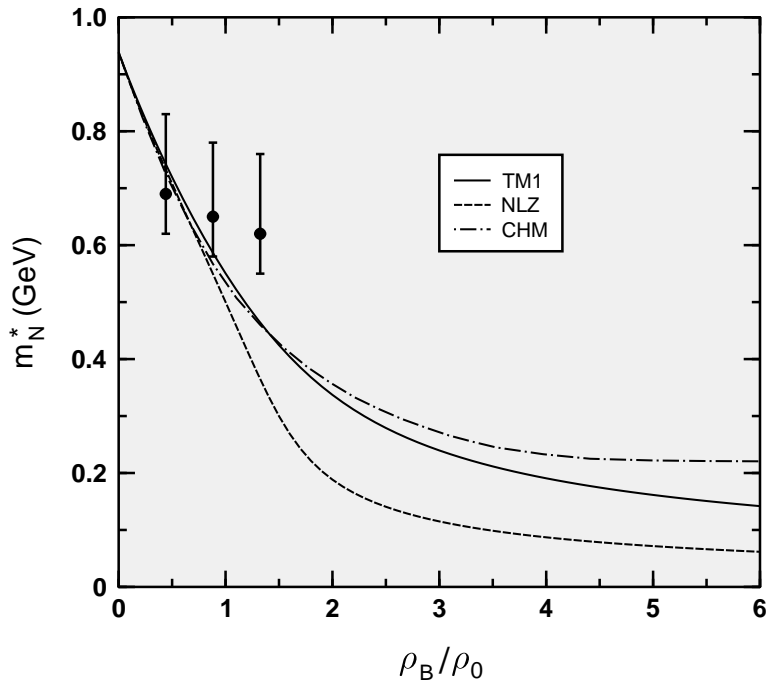}
}
\caption
{Nucleon effective mass as function of baryon density
predicted by different hadronic models. The dots show
estimates~\cite{Fur96} obtained
from the QCD sum rules.}
\label{fig6}
\end{figure}

An important characteristic of nuclear matter is the in--medium nucleon
mass~$m_N^*$\,. Experiments on inelastic electron--nucleus scattering
show that at \mbox{$\rho_B\sim\rho_0$}\,, this mass is lowered to about
$0.7 m_N$\,. By solving the gap equations (\ref{nmas}), (\ref{nmas1})
one can calculate the $\rho_B$ dependence of~$m_N^*$ within the CHM as
well as in the TM1 and NLZ models. The results are shown in
Fig.~\ref{fig6}. All the models predict a rapid drop of $m_N^*$ at
$\rho_B\lesssim 2\rho_0$ and much slower decrease at higher densities.
This behavior shows the tendency towards restoration of chiral
symmetry at high~$\rho_B$\,.

Having in mind that predictions of the NLZ model at high densities
strongly deviate from the results of other models, below we use only
the CHM, VCM, and the TM1 model for the comparison with quark models.

\noindent

\section{Deconfined phase}

\subsection{The bag model}

The simplest description of deconfined phase is given by the MIT
bag model (BM). Within this model pressure of homogeneous nonstrange
quark matter at zero temperature is expressed as
\bel{bag1}
P=P_0\hsp (\mu)-B\,,
\ee
where $\mu=\mu_B/3$ is the chemical potential of light
quarks, $B$ is the bag constant and
\bel{bag2}
P_0\hsp (\mu)=\frac{\ds N_f\mu^4}{\ds 4\pi^2}
\ee
is pressure of ideal gas of massless quarks with $N_f$ flavors.
Below we assume that only light $u,d$ quarks are present, i.e. $N_f=2$.
We consider the simplest version of the BM neglecting perturbative
corrections due to the color--magnetic interaction. Thus, the only
nonperturbative aspect of the model is associated with the bag constant $B$.
The BM has been widely used for modelling a phase transition from the
baryon--free hadron matter into the quark--gluon plasma at nonzero
temperature. The critical temperature $T_c$ for this transition
is determined not only by the bag constant but also by the numbers of
active degrees of freedom in respective phases. Reasonable values,
$T_c\sim 160$ MeV, are obtained~\cite{Kap84} with $B^{1/4}\sim 200$ MeV.
The calculations below are made for $B^{1/4}$=145, 165 and 200 MeV.

\begin{figure}
\centerline{
\includegraphics[height=9cm]{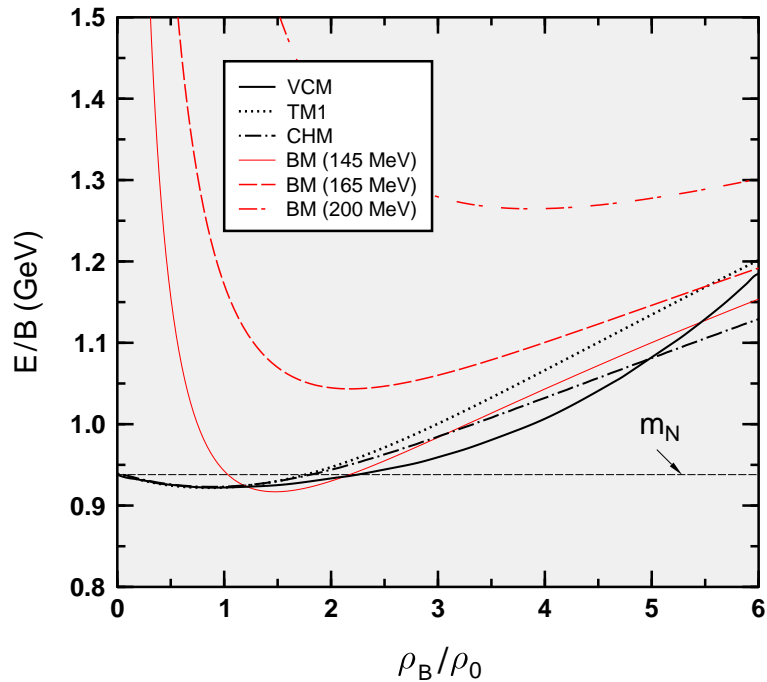}
}
\caption
{Comparison of energy per baryon of cold symmetric matter calculated
within the MIT bag and different hadronic models. Values of $B^{1/4}$
in MeV are given in parentheses.}
\label{fig7}
\end{figure}
\begin{figure}
\centerline{
\includegraphics[height=9cm]{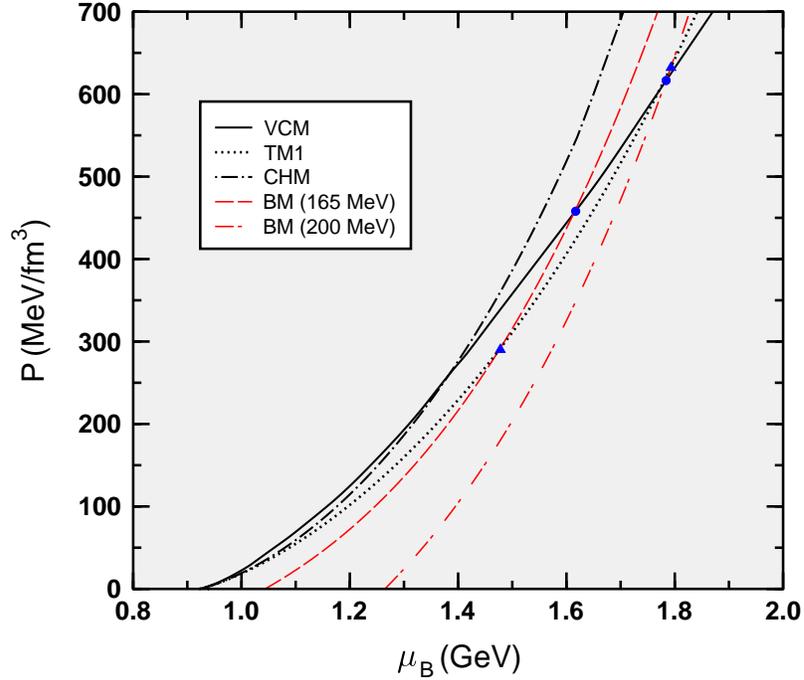}
}
\caption
{The same as in Fig.~\ref{fig7}, but for pressure vs baryon
chemical potential. Dots (triangles) mark the points of possible
phase transitions from the VCM (TM1) hadronic phase to the BM quark
phase with \mbox{$B^{1/4}=165$ and 200 MeV}.}
\label{fig8}
\end{figure}

\begin{figure}
\centerline{
\includegraphics[height=9cm]{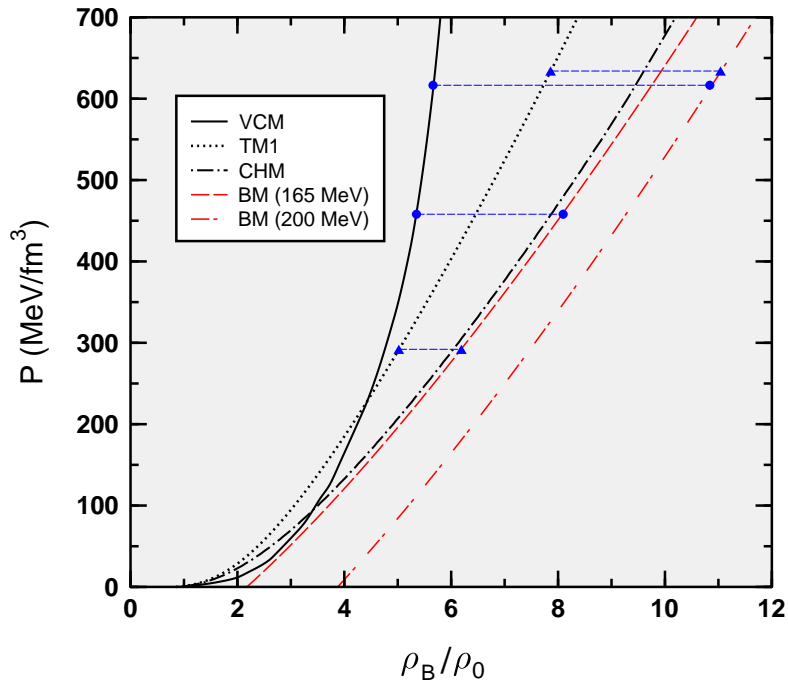}
}
\caption
{The same as in Fig.~\ref{fig7}, but for pressure vs baryon
density. Pairs of dots (triangles)
connected by dashed lines show mixed hadron--quark states
obtained by matching the EOS, predicted by the VCM (TM1 model)
and the BM with \mbox{$B^{1/4}=165$ and 200 MeV}.}
\label{fig9}
\end{figure}

In Figs.~\ref{fig7}--\ref{fig9} we compare the EOS predicted by the
hadronic models (VCM, CHM and~TM1) and the BM for different values of $B$\,.
Figure~\ref{fig7} shows energy per baryon as a function of baryon density.
It is clear that small values of bag constant, $B^{1/4}\lesssim 150$\,MeV,
are unrealistic because in this case quark matter would be more stable at $\rho_B\sim\rho_0$
than cold nuclear matter. Due to this reason, below we
consider only two values of bag constant corresponding to $B^{1/4}= 165$ and $200$\,MeV.

Figure~\ref{fig8} shows pressure as a function of chemical potential
calculated for the same models. From this figure one can see that at
\mbox{$B^{1/4}=165$ and 200 MeV} the transition from the hadronic to
the quark phase is possible only for the VCM and the TM1 model.
Filled symbols in Fig.~\ref{fig8} mark points of intersection between
respective pressure curves. By using the Gibbs rules~\mbox{(\ref{gr1})--(\ref{gr2})}
it is possible to calculate parameters of the
\begin{table}[h]
\caption{Parameters of hadron--quark phase transition predicted by
matching the hadronic models and the BM with different
bag constants $B$\,.}
\label{tab2}
\begin{center}
\begin{tabular}{|c|c|c|c|c|c|c|c|} \hline\hline
&$B^{1/4}$, MeV&$\mu_B$, GeV&$P$, MeV/fm$^3$&
$\rho_B^{(1)}/\rho_0$&$\rho_B^{(2)}/\rho_0$&$\epsilon^{(1)}$, GeV/fm$^3$&
$\epsilon^{(2)}$, GeV/fm$^3$\\
\hline
TM1&165& 1.478 & 292 & 5.02 & 6.19  & 0.97 & 1.20\\
TM1&200& 1.793 & 634 & 7.86 & 11.04 & 1.76 & 2.73\\
VCM&165& 1.617 & 458 & 5.35 & 8.10  & 1.01 & 1.77\\
VCM&200& 1.783 & 615 & 5.66 & 10.84 & 1.10 & 2.67\\
\hline\hline
\end{tabular}
\end{center}
\end{table}
quark--hadron mixed phase predicted by these models. The results of
such calculation are summarized in Table~\ref{tab2}. A more detailed
information on the resulting EOS with the hadron--quark phase transition
is given by Fig.~\ref{fig9}. However, the calculations show that the
phase transition occurs at rather large values of $\mu_B$ and $\rho_B$,
in the region where applicability of hadronic models is questionable.
%

\subsection{The massive quasiparticle model}

A more realistic model of the quark phase, the MQM, has been proposed
in Refs.~\cite{Pes96,Sch97}. The case of hot baryon--free matter has
been studied in Ref.~\cite{Pes96}. In Ref.~\cite{Sch97} a similar model
was formulated to study the EOS of quark matter at $T=0,\,\mu\neq 0$\,%
\footnote{
 It should be noted that the dependence of strong
 coupling constant on $\mu$ was disregarded in Ref.~\cite{Sch97}.
}.
Within the MQM it is assumed that cold quark matter may be regarded as
an ideal gas of quarks with nonzero effective mass $m=m\hsp (\mu)$.
Arguments in favor of this picture follow from calculations based on the
hard thermal loop resummation technique developed in
Refs.~\cite{Kar97,Bla99,And99}. In the case of hot baryon--free matter
it was possible to reproduce lattice results by using only leading
order diagrams for quark and gluon self--energies. On the other hand,
at small strong coupling constant, $\alpha_s=g^2/4\pi$\,, one obtains
results consistent with the perturbation theory up to the $\alpha_s^2$
order. The case of cold quark matter has been recently
studied~\cite{Bai00} within the hard density loop resummation
technique. It was shown that dependence of pressure on quark chemical
potential can be well approximated by treating quarks as ideal gas
of massive quasiparticles. However, the analytic formula
$P=P_{\rm{id}}(m,\mu)$ suggested in Ref.~\cite{Bai00} is
thermodynamically inconsistent in the case when
$m\hsp (\mu)\neq\textrm{const}$ (see below).

As shown in Refs.~\cite{Kli82,Wel82a,Kaj83}, at large enough 3--momenta the
dispersion relation for quarks can be interpreted in terms of
quasiparticles with nonzero effective mass $m$\,. In the leading order
in $\alpha_s$ this mass can be expressed as~\cite{Kaj83}
\bel{emas}
m=\sqrt{\frac{2\alpha_s}{3\pi}}\hsp\mu\,.
\ee
One should have in mind that at small values of $\mu$
corresponding to $\alpha_s\gtrsim 1$ (see below) the applicability
of \re{emas} becomes questionable.

In our calculations we use the three--loop expression~\cite{PDG00} for
the strong coupling constant $\alpha_s=g^2/4\pi$
as function of the renormalization scale $Q$
\bel{sdep}
\alpha_s\hsp (Q)=\frac{4\pi}{\beta_0 L}
\left\{1-\frac{2\hsp\beta_1}{\beta_0^{\hsp 2}}\frac{\log{L}}{L}+
\frac{4\beta_1^{\hsp 2}}{\beta_0^4 L^2}\left[(L-1/2)^2+
\frac{\beta_0\beta_2}{8\beta_1^{\hsp 2}}-\frac{5}{4}\right]\right\}\,.
\ee
Here $L=\log{(Q^{\hsp 2}/\Lambda^2)}$ and
\begin{eqnarray}
&&\beta_0=11-2\hsp N_f/3,\,\,\beta_1=51-19\hsp N_f/3,\nonumber\\
&&\beta_2=2857-5033\hsp N_f/9+325\hsp N_f^{\hsp 2}/27\,.\label{coeff}
\end{eqnarray}
The cutoff momentum $\Lambda$ is fixed by
the condition $\alpha_s\,(2\,{\rm GeV})=0.3089$~\cite{PDG00},
which gives \mbox{$\Lambda\simeq 0.4178$}\,GeV for $N_f=2$.
From dimensionality arguments it is clear that $Q\sim\mu$\,.
As in Refs.~\cite{Sch97,Bai00,Fra01}, it is assumed that
$Q=\gamma\hsp\mu$ where the coefficient $\gamma$ is of the
order of unity and does not depend on $\mu$\,. At fixed $\gamma$
physical values of $\mu$ correspond to $Q>\Lambda$\,.

\begin{figure}
\centerline{
\includegraphics[height=9cm]{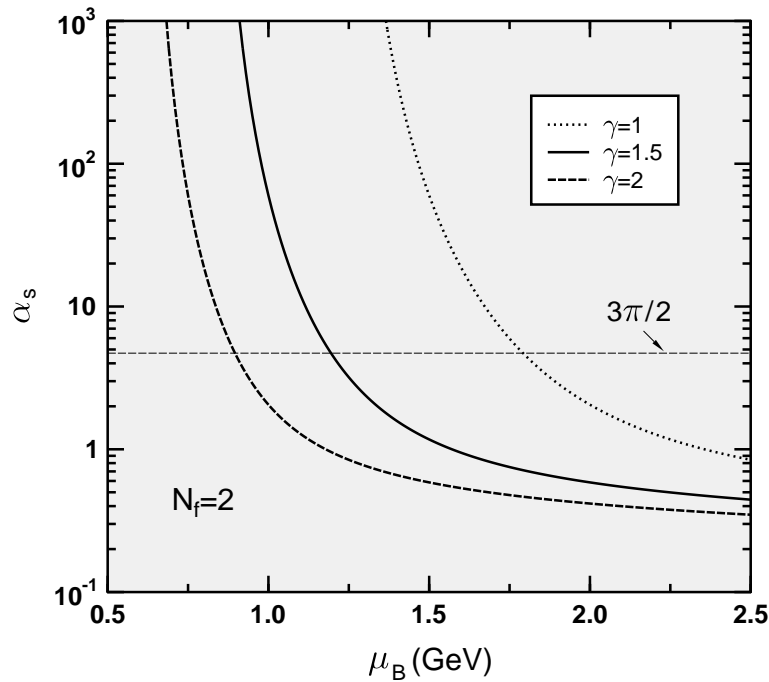}
}
\caption
{The QCD coupling constant as function of baryon chemical
potential at $N_f=2$ and different values of the
parameter $\gamma$\,. Thin dashed line shows maximal
possible value of $\alpha_s$ within the MQM.}
\label{fig10}
\end{figure}
Further it is postulated that the density of quark matter
equals to the density of ideal gas of massive fermions
\bel{qdni}
\rho=3\rho_B=\rho_{\rm{id}}\hsp(\mu)\equiv
\frac{N_f}{\pi^2}\left(\mu^2-m^2\right)^{3/2}\Theta (\mu-m)\,,
\ee
where $\Theta (x)\equiv (1+{\rm sgn}\hsp x)/2$\,. Thus, the quark
density vanishes at $\mu\leq m$. This condition is satisfied at
$\mu\leq\mu_c$\, where $\mu_c$ is the critical chemical potential
determined from \re{emas} with $m=\mu$\,. The latter condition is equivalent to
$\alpha_s=3\pi/2$\,.
The calculation shows that $\mu_c\simeq 0.5974/\gamma$~GeV for
$N_f=2$\,.

\begin{figure}
\vspace*{1cm}
\centerline{
\includegraphics[height=9cm]{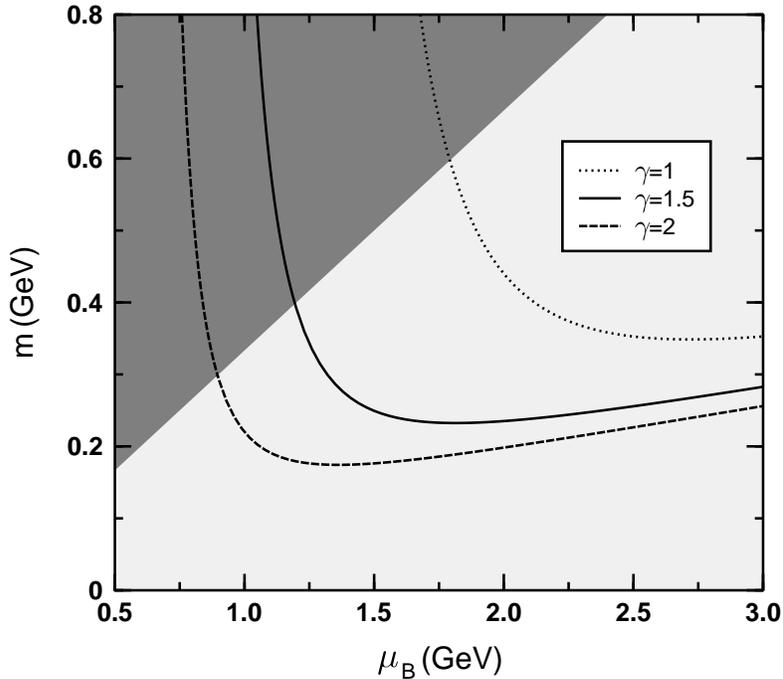}
}
\caption
{Effective mass of $u,d$ quarks $m$ within the MQM vs
baryon potential of quark matter~$\mu_B$. Unphysical region
is shown by shading.}
\label{fig11}
\end{figure}

Figs.~\ref{fig10}--\ref{fig11} show $\alpha_s$ and $m$ as functions of
baryon chemical potential $\mu_B=3\mu$ for several values of the parameter
$\gamma$. The shaded region in Fig.~\ref{fig11} corresponds to the
domain where $m>\mu$\,. It is seen that quark masses are very sensitive
to the choice of~$\gamma$. At large $\mu_B$ quark masses increase,
although slower than $\mu_B$\,. As discussed in Ref.~\cite{Wel82b}, the
existence of nonzero quark mass does not contradict to the restoration of
chiral symmetry at large~$\mu_B$\,.

\begin{figure}
\centerline{
\includegraphics[height=9cm]{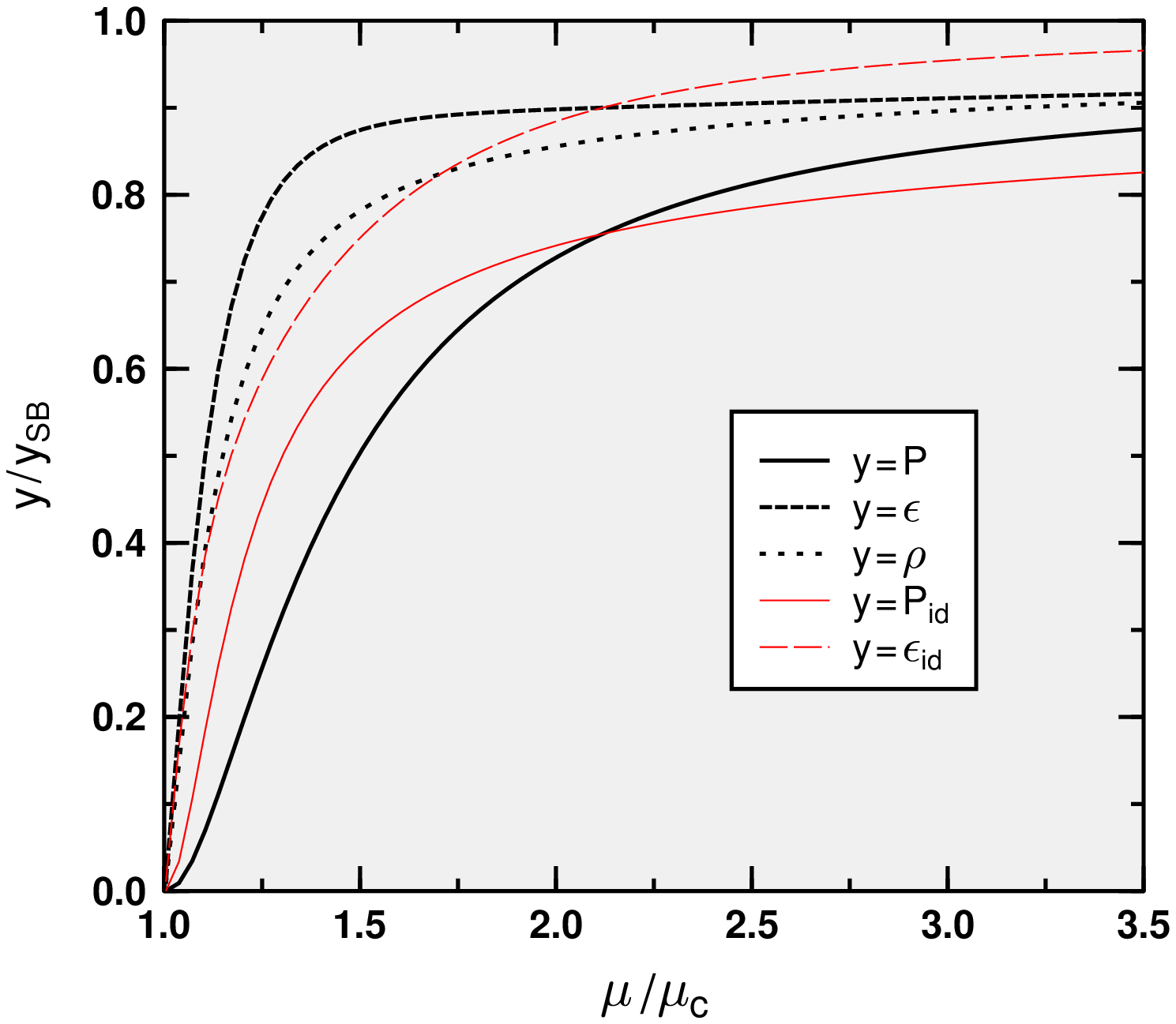}
}
\caption
{Pressure $P$, energy density $\epsilon$ and number density $\rho$
of quarks within the MQM (normalized to their
Stephan--Boltzmann limits, $m\to 0$) as functions of quark chemical
potential $\mu$ in units of its critical value $\mu_c$\,. Thin solid and
dashed lines show pressure and energy density calculated by using the
ideal gas approximation $B\hsp (\mu)=0$.}
\label{fig12}
\end{figure}

Using \re{qdni} and thermodynamic identity (\ref{dens})
one can calculate pressure as a function of~$\mu$\,:
\bel{pre2}
P=\int\limits_{\mu_c}^\mu \textrm{d}\mu_1 \rho_{\rm{id}}(\mu_1)\,.
\ee
Here we assume that $P(\mu_c)=0$\,, i.e. pressure is zero at vanishing
quark density. If one would replace $m\hsp (\mu_1)$ in the integrand
$\rho_{\rm{id}}$ by the ''constant'' mass $m\hsp (\mu)$\,, one would
obtain the well--known expression~\cite{Bai00} for pressure of ideal
gas $P_{\rm{id}}$. However, as shown in Ref.~\cite{Gor95}, the
approximation \mbox{$P=P_{\rm{id}}$} is
thermodynamically inconsistent when masses depend on $T$ or~$\mu_B$\,.
In~particular, \re{dens} does not hold in this case. The
thermodynamic consistency can be recovered by introducing the effective
bag constant $B(\mu)=P_{\rm{id}}-P$\,.

Figure~\ref{fig12} shows thermodynamic quantities calculated within the
MQM by using Eqs.~(\ref{qdni})--(\ref{pre2}), (\ref{enden}).
Density, pressure and energy density are shown as functions of the
dimensionless variable $\mu/\mu_c$\,. All quantities are given as ratios
to their respective limits for massless quarks~($P_{SB}=\epsilon_{SB}/3=P_0\hsp (\mu)$).
In this representation the results do not depend on $\gamma$\,. In Fig.~\ref{fig12}
we also demonstrate inaccuracy of the approximation $P=P_{\rm{id}}$ used in
Ref.~\cite{Bai00}. It is seen that $B(\mu)$ is typically of order $0.1 P_{\rm{id}}$
and changes sign at $\mu\sim 2\hsp\mu_c$\,.

\begin{figure}
\centerline{
\includegraphics[height=9cm]{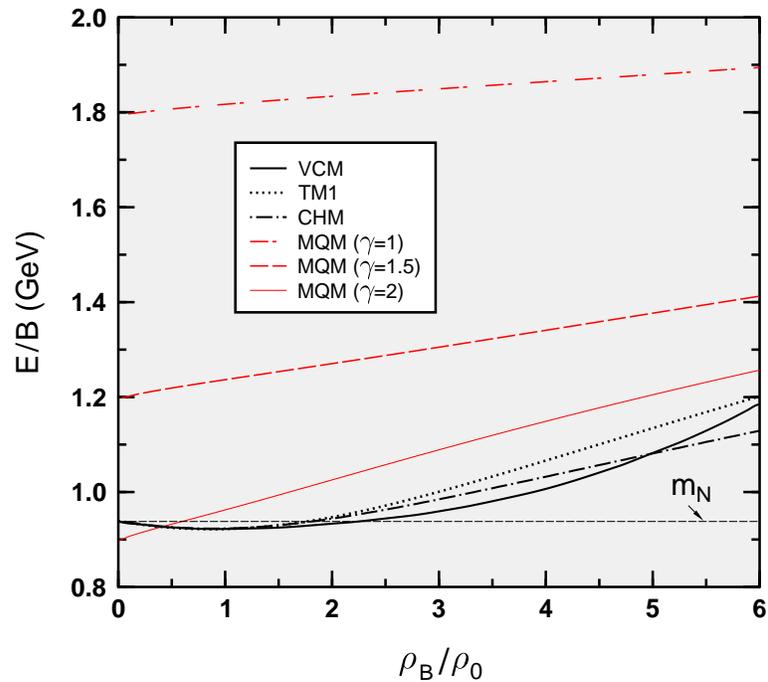}
}
\caption
{Comparison of energy per baryon of cold symmetric matter calculated
within the MQM and different hadronic models.}
\label{fig13}
\end{figure}

\begin{figure}
\centerline{
\includegraphics[height=9cm]{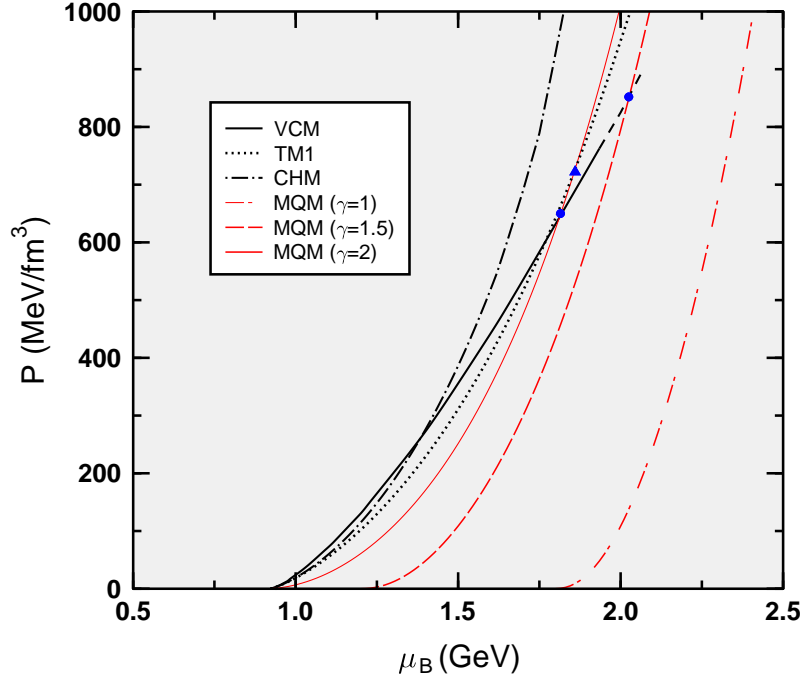}
}
\caption
{The same as in Fig.~\ref{fig13}, but for pressure vs
baryon chemical potential. Filled symbols show points of
possible phase transitions.}
\label{fig14}
\end{figure}

Energy per baryon and pressure calculated within the MQM are compared with
predictions of hadronic models in Figs.~\ref{fig13}--\ref{fig14}.
Unlike the BM, in this model the energy per baryon
increases monotonically with $\rho_B$, even at low densities. According to
Fig.~\ref{fig13}, the MQM with~$\gamma>2$ is
unrealistic since in this case the quark matter would be energetically
more favorable at $\rho_B\lesssim\rho_0$ than normal nuclear matter.

According to Fig.~\ref{fig14} at $1.5\le\gamma\le 2$ the MQM pressure
curve intersects only with curves predicted by the VCM and the
TM1 model. Parameters of possible phase transitions are given in
Table~\ref{tab3}. One can see that quark phase appears only at rather
high baryonic densities, namely, at
$\rho_B>\rho^{(1)}_B\gtrsim 6\hsp\rho_0$\,.  At
$\gamma\lesssim 1$ there are no intersections of
hadronic and quark branches, at least at not too large $\mu_B$.
\begin{table}[h]
\caption{Parameters of quark--hadron phase transition predicted by
matching the~VCM, TM1~model, and MQM  at different $\gamma$\,.}
\vspace*{3mm}
\label{tab3}
\begin{center}
\begin{tabular}{|c|c|c|c|c|c|c|} \hline\hline
$\gamma$~~~~~&$\mu_B$, GeV&$P$, MeV/fm$^3$&
$\rho_B^{(1)}/\rho_0$&$\rho_B^{(2)}/\rho_0$&$\epsilon^{(1)}$, GeV/fm$^3$&
$\epsilon^{(2)}$, GeV/fm$^3$\\
\hline
1.5 (\textrm{VCM}) & 2.025 & 852 & 6.0 & 13.2 & 1.21 & 3.71\\
2.0 (\textrm{VCM}) & 1.815 & 650 & 5.7 & 9.1  & 1.11 & 2.17\\
2.0 (\textrm{TM1}) & 1.860 & 725 & 8.6 & 10.6 & 1.95 & 2.67\\
\hline\hline
\end{tabular}
\end{center}
\end{table}

We have also checked sensitivity of the results to the choice of $N_f$
in the QCD coupling constant. The calculation has been made using
$N_f=3$ in Eqs.~(\ref{sdep})--(\ref{coeff}), but retaining $N_f=2$
(only $u$ and $d$ quarks) in thermodynamic quantities. This leads to
a slight increase of $P(\mu_B)$ that lowers the phase transition
density by not more than 10\%.

\subsection{The Nambu--Jona-Lasinio model}

The NJL model~\cite{Nam61,Lar61} is one of the most popular models dealing with
constituent quarks. There are several advantages of the NJL model
compared to other models considered above.  First, it respects chiral
symmetry of strong interactions, second, it explicitly takes into account
negative energy (Dirac sea) states, and third, it describes well meson
phenomenology.
Different versions of this model have been extensively used to describe
the EOS of equilibrium and nonequilibrium quark matter at finite $T$ and
$\mu_B$\,.

Here we use results of our calculations~\cite{Mis99,Mis00,Mis01} within
the SU(3)--flavor version of the model suggested in Ref.~\cite{Reh96},
but with an additional term due to vector--axial-vector interaction.
The color-singlet part of the Lagrangian in the mean field approximation
can be written as
\begin{eqnarray}
{\cal L}&=&\sum_f\psib_f\,(i\dd-m_f-\gamma_0\,G_V\rho_{\mbs{Vf}})\,\psi_f
-\frac{G_S}{2}\sum_f\rho_{\mbs{Sf}}^{\,2}\nonumber\\
&+&\frac{G_V}{2}\sum_f\rho_{\mbs{Vf}}^{\,2}
+4K\prod_f\rho_{\mbs{Sf}}\,.\label{lagrm}
\end{eqnarray}
Here~$\psi_f$~is~the~field operator of quarks with
flavor \mbox{$f=u,d,s$} and
\begin{eqnarray}
\rho_{\mbs{Sf}}&=&<\psib_f\psi_f>\,,\label{denss}\\
\rho_{\mbs{Vf}}&=&<\psib_f\gamma_0\psi_f>\label{denv}
\end{eqnarray}
are their scalar and vector densities.
Angular brackets in Eqs.~(\ref{denss})--(\ref{denv})
denote quantum--statistical averaging.
$G_S$, $G_V$ and $K$ in \re{lagrm} are, respectively, the coupling
constants of scalar, vector and flavor--mixing interactions.

The constituent quark masses, $m_f$, are determined from the coupled set of
gap equations
\bel{gape}
m_f=m_{0f}-G_S\,\rho_{\mbs{Sf}}+
2K\,\prod_{f'\neq f}\rho_{\mbs{Sf'}}\,,
\ee
where $m_{0f}$ is the bare (current) mass of quarks with
flavor $f$\,.

The NJL model is an effective, non--renormalizable model.  To
regularize the divergent contribution of negative energy states of
the Dirac sea, one must introduce an ultraviolet
cut--off. Below the 3--momentum cut--off $\Theta(\Lambda-p)$ is used
in divergent integrals. The model parameters $m_{0f}, G_S, K, \Lambda$
can be fixed by reproducing the observed masses of $\pi, K$\,, and
$\eta'$ mesons as well as the pion decay constant. As shown in
Ref.~\cite{Reh96}, a reasonable fit is achieved with the following
input parameters:
\begin{eqnarray}
m_{0u}=m_{0d}&=&5.5~{\rm MeV},~~~m_{0s}=140.7~{\rm MeV},\\
G_S=20.23~{\rm GeV}^{-2},&&\Lambda=0.6023~{\rm GeV},
~~~~K=155.9~{\rm GeV}^{-5}.
\label{para}
\end{eqnarray}

In principle, the vector coupling constant may be
extracted by fitting the nucleon axial charge or masses
of vector mesons. It was shown~\cite{Kli90} that the ratio
$\xi=G_V/G_S$ should be of the order of unity.
However, as discussed in Ref.~\cite{Han01}, the accuracy
of such fitting procedure is rather low. Due to uncertainty in the
parameter $G_V$\,, below we present results for various values of
$\xi$ from the interval $0\le\xi\le 1$\,.

Let us consider cold isospin--symmetric nonstrange matter with baryon
density $\rho_B$\,. In this case $\rho_{\mbs{Vs}}=0,
\rho_{\mbs{Vu}}=\rho_{\mbs{Vd}}=3\rho_B/2$\,. The energy density
$\epsilon$ can be calculated directly from the
Lagrangian~(\ref{lagrm}) with the result~\cite{Mis00}
\bel{ende1}
\epsilon=\sum_f
\left[\hsp\epsilon_{\mbs{Kf}}+\frac{G_S}{2}
\rho_{\mbs{Sf}}^{\hsp 2} +\frac{G_V}{2}
\rho_{\mbs{Vf}}^{\hsp 2}\right]
-4K\prod_f\rho_{\mbs{Sf}}+\epsilon_0\,.
\ee
Here  $\epsilon_{\mbs{Kf}}$ is the kinetic term
which includes also ''active'' negative energy states
with momenta~$p<\Lambda$:
\bel{kend}
\epsilon_{\mbs{Kf}}=\frac{3}{\pi^2}\int
\limits_\Lambda^{\hspace*{1ex}p_{\scriptscriptstyle{Ff}}}
\textrm{d}p\,p^{\hsp 2}\sqrt{m_f^{\,2}+p^{\,2}}\,,
\ee
where $p_{\mbs{Ff}}=(\pi^2\rho_{\mbs{Vf}}/3)^{1/3}$ is the
$f$--quark Fermi--momentum. Scalar densities can be calculated
by using the relation
\bel{sden1}
\rho_{\mbs{Sf}}=\left(\frac{\ds\partial\hsp\epsilon_{\mbs{Kf}}}
{\ds\partial\hsp m_f}
\right)_{p_{\scriptscriptstyle{Ff}}}\,.
\ee
The constant $\epsilon_0$ in the l.h.s. of \re{ende1}
is introduced in order to set the energy density of the
physical vacuum equal to zero. This constant
can be calculated by solving the gap equations~(\ref{gape})
for the vacuum case, $\rho_B=0$\,. Finally, pressure can be
found by using thermodynamic identities $\mu_B=d\epsilon/d\rho_B$
and $P=\mu_B\rho_B-\epsilon$\,.

\begin{figure}
\centerline{
\includegraphics[height=9cm]{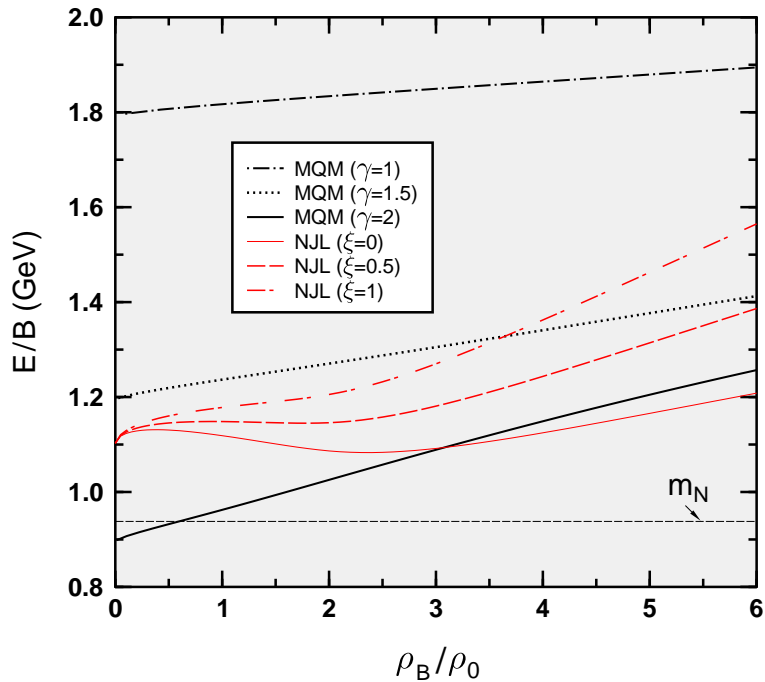}
}
\caption
{Comparison of energy per baryon of cold symmetric matter calculated
within the MQM and NJL model.}
\label{fig15}
\end{figure}
\begin{figure}
\centerline{
\includegraphics[height=9cm]{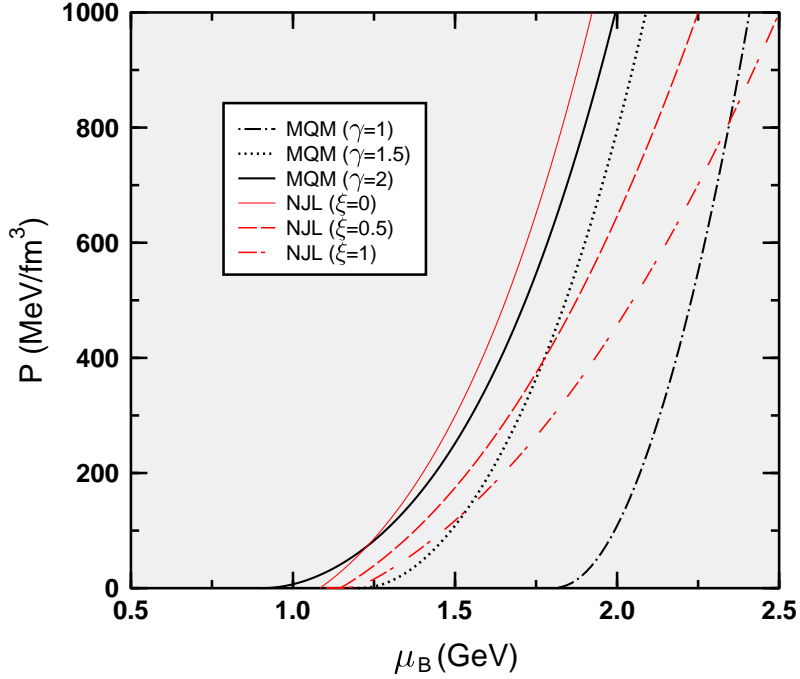}
}
\caption
{The same as in Fig.~\ref{fig15}, but for pressure vs
baryon chemical potential.}
\label{fig16}
\end{figure}
\begin{figure}
\centerline{
\includegraphics[height=9cm]{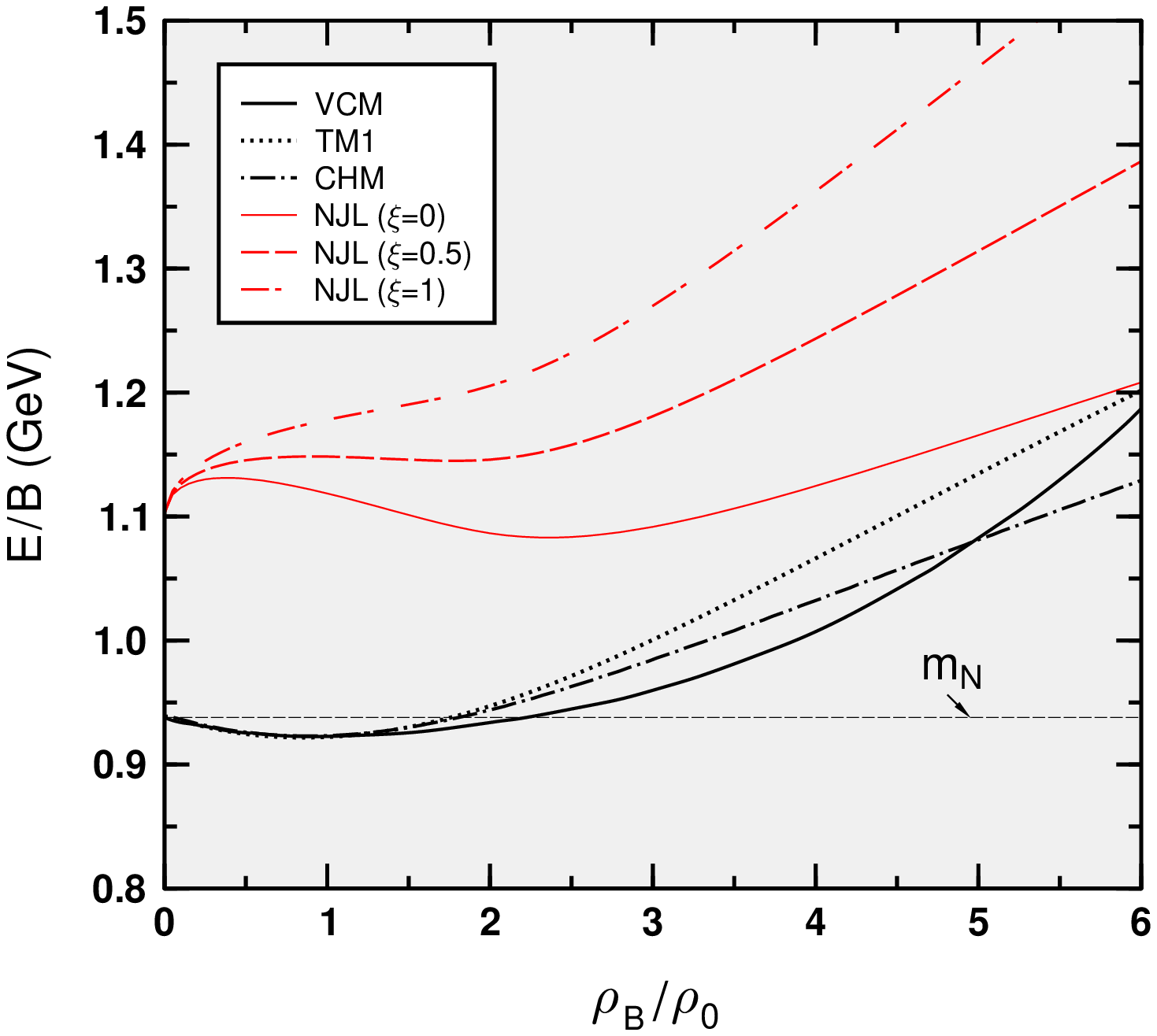}
}
\caption
{Comparison of energy per baryon of cold symmetric matter calculated
within the NJL model and different hadronic models.}
\label{fig17}
\end{figure}
\begin{figure}
\centerline{
\includegraphics[height=9cm]{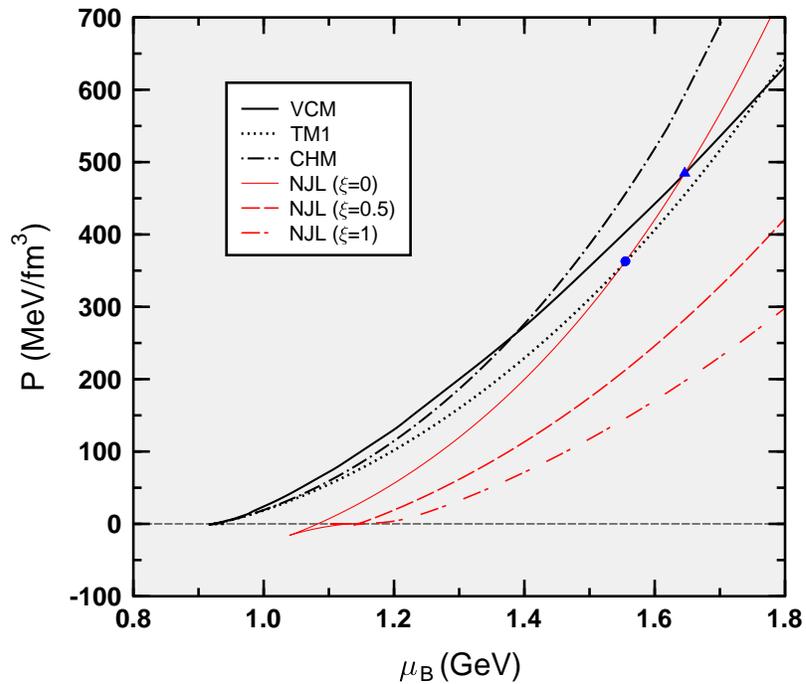}
}
\caption
{The same as in Fig.~\ref{fig17}, but for pressure vs
baryon chemical potential. Filled symbols show points of
possible phase transitions.}
\label{fig18}
\end{figure}

The results of calculations are shown in
Figs.~\ref{fig15}--\ref{fig18} for three values (0, 0.5, 1.0) of the
parameter $\xi$\,. Figs.~\ref{fig15}--\ref{fig16} show the
comparison of the MQM and NJL model. It is seen that the discrepancy
between their results is especially large in the case $\gamma=1$\,.
Unlike the MQM, the NJL model predicts metastable states
of quark matter corresponding to local minima in $E/B$
as function of $\rho_B$\,.
However, such states exist only at $\xi\lesssim 0.1$\,%
\footnote{
 As shown in Refs.~\cite{Mis00,Mis01}, they appear
 in strange matter even at larger $\xi$.
}.

In Figs.~\ref{fig17}--\ref{fig18} we compare EOS
predicted by the NJL and hadronic models. As seen in
Fig.~\ref{fig18} the hadron--quark phase transition may take place
only at small~$\xi$ which, most likely, are
not realistic. For larger $\xi$ the repulsive vector interaction
makes the NJL phase too stiff to cross any of the hadronic curves.
The phase transition parameters for $\xi=0$ are given
in Table~\ref{tab4}. Again the transition to quark matter is
possible only from the TM1 and VCM hadronic phases and predicted
critical densities are above $5\rho_0$.

At $\xi<0.71$ the pressure
isotherms calculated within the NJL model contain unstable parts
in the region $\mu_B\simeq 1-1.2$ GeV~\cite{Mis00}\,. This means
that this model itself predicts the first-order chiral phase transition in
the quark matter. However, corresponding parts of pressure isotherms lie
below the hadronic curves. Therefore, such a phase transition is
not observable due to the hadronization of quark phase.
This situation is quite general: many phase transitions found within
different quark models are predicted in regions of the $\mu_B-P$
plane where quark phase is unstable with respect to hadronization.
\begin{table}[h]
\caption{Parameters of hadron--quark phase transition
matching the~VCM and TM1 model with the NJL model ($\xi=0$).}
\vspace*{3mm}
\label{tab4}
\begin{center}
\begin{tabular}{|c|c|c|c|c|c|c|} \hline\hline
&$\mu_B$, GeV&$P$, MeV/fm$^3$&
$\rho_B^{(1)}/\rho_0$&$\rho_B^{(2)}/\rho_0$&$\epsilon^{(1)}$, GeV/fm$^3$&
$\epsilon^{(2)}$, GeV/fm$^3$\\
\hline
\textrm{TM1--NJL} & 1.555 & 363 & 5.6 & 7.2 & 1.18 & 1.54\\
\textrm{VCM--NJL} & 1.646 & 487 & 5.4 & 8.5 & 1.02 & 1.89\\
\hline\hline
\end{tabular}
\end{center}
\end{table}

\section{Conclusions and outlook}

Our goal in this paper was to find possible phase transition between
hadronic and quark phases of strongly interacting matter at high baryon
density. For each of the phases we have selected several models which
are widely discussed in
the literature. Namely, we have considered the CHM, VCM, and RFM models
for the hadronic phase, and the BM, MQM, and NJL model for the quark
phase. When necessary, we have varied model parameters in reasonable
limits to determine possible range of uncertainty. For each model we
have calculated main thermodynamic quantities, in particular, pressure
as a function of baryon chemical potential, $P(\mu_B)$\,.
Possibility of a deconfinement phase transition is signalled by an intersection
of $P(\mu_B)$ curves corresponding to the hadronic and quark phases.

A pairwise comparison of different hadronic and quark models has
shown that such an intersection is not at all a general feature but
rather an exception. For instance, the CHM, which is probably the most
advanced hadronic model at present, does not predict a
phase transition with any of quark models. Also, no phase transition is
predicted when most realistic parameters are used in quark models
($B^{1/4}\simeq 200$ MeV in the BM, $\gamma\simeq 1$ in the MQM,
or $\xi\gtrsim 0.5$ in the NJL model). Even in the cases
when a phase transition is possible, e.g. between the TM1 model and
the BM, or between the VCM and BM, characteristics of mixed phase
are very sensitive to model parameters. Moreover, this phase
transition is predicted at such a high density, above $5\rho_0$\,
where predictions of hadronic models are very unreliable.
The situation with quark models is even worse. First, it is
unclear at all how far down in density one can use these models.
Second, the predictions of various quark models differ significantly
in the region of moderately high baryon densities of interest here.
Of course, due to the asymptotic freedom of QCD
the quark phase must approach asymptotically the ideal gas limit.
However, it is unclear at present when such asymptotic behavior sets in.

In this paper we have considered only isospin--symmetric
matter, but we think that similar conclusions can be made
also for $\beta$--equilibrium neutron star matter. For example, the
comparison of EOS predicted by the CHM and NJL model (for details see
Ref.~\cite{Han01}) does not show any phase transition between hadronic
and quark phases.

As has been demonstrated in this paper, some phase transitions found in the QCD
motivated models, e.g. the chiral transition within the NJL model, occupy
regions of the $\mu_B-P$ plane where quark phases are
unstable with respect to hadronization. We expect that similar
situation takes place also for color-superconducting phases of quark
matter~\cite{Alf98,Sch99}, although this question deserves a special
study.

We should conclude that after more than 30 years of model
building in both the hadronic and quark sectors the situation
regarding a hadron--quark phase transition and EOS at high baryon
densities remains rather uncertain. We believe that progress in this
field can be achieved by developing new class of models where
both the hadronic and quark degrees of freedom are treated
within a unified theoretical framework.
Attempts to construct such a model were made in Refs.~\cite{Nik99,Sch00}.
Such unified approach should
include constraints from nuclear physics (existence of the nuclear
bound state) and QCD (chiral symmetry, asymptotic freedom).
It is quite possible that the transition from hadronic to
quark degrees of freedom will be continuous~\cite{Mis00}, like
ionization in atomic systems~\cite{Lan80}. Indications of such behavior
are found in recent lattice calculations~\cite{Eng99}. All this means
that the problem of hadron--quark transition at high baryon densities
remains a challenge for theorists.

\section*{Acknowledgments}

The authors thank P.J. Ellis and D.H. Rischke for fruitful discussions.
I.N.M. and L.M.S. acknowledge financial support from DAAD and GSI, Germany.
This work has been partially supported by the RFBR Grant No.~00--15--96590.

\end{document}